# Spin-Flipping in Pt and at Co/Pt Interfaces

H.Y.T. Nguyen, W.P. Pratt Jr., J. Bass

Department of Physics and Astronomy, Michigan State University, East Lansing, MI USA

## ABSTRACT

There has been recent controversy about the magnitude of spin-flipping in the heavy metal Pt, characterized by the spin-diffusion length $l_{sf}^{Pt}$. We propose a resolution of this controversy, and also present evidence for the importance of a phenomenon neglected in prior studies of transport across sputtered Ferromagnet/Pt (F/Pt) interfaces, spin-flipping at the interface. The latter is characterized by an interface spin-flipping parameter, $\delta_{Co/Pt}$, that specifies the probability $P = [1 - \exp(-\delta)]$ of a conduction electron flipping its spin direction as it traverses a Co/Pt interface. From studies of the Current-Perpendicular-to-Plane (CPP) Resistances and Magnetoresistances of sputtered ferromagnetically coupled Co/Pt multilayers by themselves, and embedded within Py-based Double Exchange-biased Spin-Valves, we derive values at 4.2K of $\delta_{Co/Pt} = 0.9^{+0.5}_{-0.2}$, interface specific resistance, $AR^*_{Co/Pt} = 0.74 \pm 0.15$ fΩm$^2$, and interface spin-scattering asymmetry, $\gamma_{Co/Pt} = 0.53 \pm 0.12$. This value of $\delta_{Co/Pt}$ is much larger than ones previously found for five other interfaces involving Co but not Pt. To derive $\delta$ requires knowledge of $l_{sf}^{Pt}$ for our sputtered Pt, which we obtain from separate measurements. Combining our results with those from others, we find that $l_{sf}^{Pt}$ for Pt is approximately proportional to the inverse resistivity, $1/\rho_{Pt}$.

## 1. Introduction

The large atomic number of Platinum (Pt) leads to expectation of a large spin-orbit interaction that should produce strong flipping of conduction electron spins as the electrons traverse a thin Pt layer. There has recently been controversy about the rate of spin-flipping in Pt, with values of the spin-diffusion length, $l_{sf}^{Pt}$, reported to range from 0.5 nm to 14 nm [1-7]. In this paper we propose an explanation for the range of values and then extend studies of spin-flipping involving Pt to Co/Pt interfaces, where we find a spin-flipping parameter, $\delta_{Co/Pt} = 0.9^{+0.5}_{-0.2}$, much larger than those found for ferromagnetic/non-magnetic (F/N) or (F1/F2) interfaces involving Co but not Pt [8-12]. This large value for Co/Pt implies that a polarized current likely strongly degrades in crossing any F/Pt interface, a phenomenon neglected in prior studies of transport in F/Pt multilayers (see [13]).

Few of the authors of prior papers on $l_{sf}^{Pt}$ intimated that the spin-diffusion length in a nominally pure metal such as Pt might not be intrinsic. But it is not. Rather, it is largely determined by the (mostly unknown) impurities and defects that scatter conduction electrons in the metal. Even at room temperature (295K), where scattering by phonons is large, the resistivity of a sputtered metallic layer due to scattering from defects and unknown impurities is typically comparable to that due to phonons. At cryogenic temperatures, where phonon scattering is negligible, scattering from defects and unknown impurities dominates, and the spin-diffusion length is no more intrinsic than the residual resistivity. For these reasons, it is inappropriate to directly compare values of $l_{sf}^{Pt}$ derived for samples of different purity and measured at different temperatures. In the absence of more detailed knowledge, one could try to compare values of $l_{sf}^{Pt}$ at the same value of the resistivity (taken as a rough measure of the total scattering in the layer). Fig. 1 shows a plot of reported values of $l_{sf}^{Pt}$ plotted against the independently measured inverse resistivity ($1/\rho_{Pt}$)--or in one case (see caption) by our best estimate of this inverse resistivity as the sum of a typical residual resistivity and the expected phonon resistivity. The straight line in Fig. 1, required to pass through zero, is a least squares fit to the four data points with the largest values of ($1/\rho_{Pt}$), weighted equally (i.e., not taking account of their specified uncertainties). Fig. 1 shows that most of the data scatter roughly around this line. We take this behavior as evidence that $l_{sf}^{Pt}$ in Pt is approximately proportional to $1/\rho_{Pt}$, and that the value of $l_{sf}^{Pt}$ used in our present analysis below (second point from the right, see section 2 below) is reasonable [Note: our derived value of $\delta_{Co/Pt}$ is insensitive to values of our new $l_{sf}^{Pt}$ spanning its range of uncertainty]. Presumably

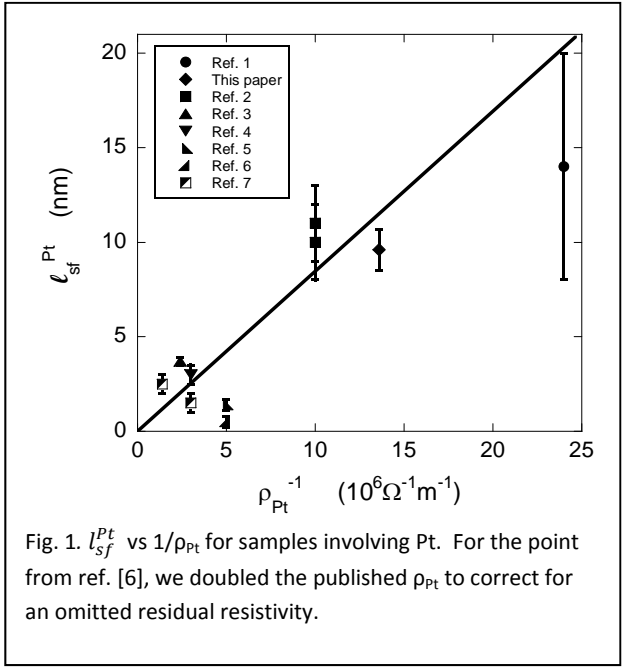

Fig. 1. $l_{sf}^{Pt}$ vs $1/\rho_{Pt}$ for samples involving Pt. For the point from ref. [6], we doubled the published $\rho_{Pt}$ to correct for an omitted residual resistivity.

the different deviations from the line indicate differences in details of impurity contents.

In Fig. 1, the four points to the right were all measured at cryogenic temperatures, whereas the six points to the left were all measured at room temperature. We take the fact that both sets of data scatter about the same line as indicating that the rate of spin-flipping due to phonon scattering, which is important at room temperature, is in the same ballpark as that due to impurity scattering, which is dominant at low temperature

The last issue is whether $l_{sf}^{Pt}$ in Pt might be affected by contact with an F-layer. A study by Lim et al. [14] suggests that Pt in contact with Permalloy (Py = Ni$_{0.8}$Fe$_{0.2}$) becomes magnetized near the Py/Pt interface over a length scale ξ increasing from about 0.2 nm at 300K to about 0.7 nm at cryogenic temperatures. Pt as an impurity in Ni, the dominant metal in Py, stays magnetic at room temperature to concentration x ~ 0.4 [15] and to an extrapolated x ~ 0.65 at 4.2K, our measuring temperature. In contrast, Pt stays magnetic in our F-metal Co to x ~ 0.9 at room temperature and by extrapolation to still larger x at 4.2K [15]. Roughness on the atomic scale of our sputtered Co/Pt interfaces will give Pt atoms near the interface more Co neighbors that those for Pt atoms at a perfect Co/Pt interface. We thus expect greater magnetism in Pt atoms near the interface than for Ni/Pt, and thus a larger effective ξ. It might even be that all of the Pt atoms in our thin (1.1 nm) layers are magnetized at 4.2K, thereby strengthening the ferromagnetic coupling between Co layers that we need for our experiments below. If the magnetization within the Pt displays some disorder, it would add to any spin-flipping otherwise present at the interface. Since the ξ found for Py/Pt is comparable to the expected interface thickness in sputtered F/N multilayers (~ 0.6-0.8 nm) [16], it is not unreasonable to subsume any such effect into the spin-flipping due to the interface, which we do.

With this background about $l_{sf}^{Pt}$ in hand, we now turn to our study of $\delta_{Co/Pt}$, which determines the probability of spin-flipping, $P = [1 - \exp(-\delta)]$, as conduction electrons flowing perpendicular to a Co/Pt interface (Current-Perpendicular-to-Plane (CPP) geometry) cross that interface. Recently published values of $\delta_{F/N}$ or $\delta_{F1/F2}$ for sputtered Co/Cu [8], Co$_{91}$Fe$_9$/Cu [9], Co/Ni [10], Co/Ru [11], and Co/Ag [12] range from about 0.2 to 0.35, indicating modest spin-flipping in all five cases. In this paper we present evidence that the parameter for the sputtered Co/Pt interface, $\delta_{Co/Pt} = 0.9^{+0.5}_{-0.2}$, is much larger than the other five cases, consistent with the large spin-orbit interaction expected for Pt as noted above.

Our analysis involves applying the theory of Valet and Fert (VF) [17] to resistance and magnetoresistance data on magnetic multilayers measured in the CPP geometry. For a general multilayer, the VF theory must be applied numerically, matching boundary conditions at interfaces as described in refs. [17,18]. In the present study, we assume that the VF parameters for all metals and interfaces in our samples, except for Pt and the Co/Pt interfaces, are fixed by prior experiments made in our laboratory. Regular cross-checks, and internal consistency of repeated data sets, strongly suggest that these parameters are reproducible in our laboratory to within their specified uncertainties. We will describe below how we obtain the parameters for Pt and for Co/Pt interfaces.

2. Samples.

2a. Multilayer Fabrication and Structures.

The CPP-MR measurements in this paper were made at 4.2K using the crossed-superconductor technique with 150 nm thick and 1.1 mm wide superconducting Nb strips sputtered above and below the multilayer of interest [19,20]. The multilayer samples were sputtered using a system with six targets as described elsewhere [19]. With the exception of Pt, the targets and sputtering rates were the same as in previous studies of $\delta_{F/N}$ [8,12]. By the end of the last prior study with Pt, the 2.25" diam. Pt target had become so thin at its center that we feared that the next run would burn it through. For the present study, that target was cut into three pieces, and the two thicker ones were placed one above the other in a 1" diam. 'gun'. As the sputtering rate with the smaller target was slower (~ 0.19 nm/sec) than that with the larger target (~ 0.35 nm/sec), we expected the resistivity of newly sputtered 60 and 200 nm thick Pt films to be larger than those previously found with the larger gun. Indeed, from Van De Pauw measurements we estimate $\rho_{Pt}$ = 75 ± 10 nΩm, compared to 42 ± 6 nΩm for samples sputtered with the large target [1]. Since knowing the spin-diffusion length for Pt is important for our analysis, we remeasured $l_{sf}^{Pt}$ for the newly sputtered Pt using the same technique as in [1], obtaining $l_{sf}^{Pt}$ = 9.6 ± 1.1 nm. The ratio 0.7 ± 0.4 of this new value to the old value of $l_{sf}^{Pt}$ = 14 ± 6 nm overlaps with the ratio 0.6 ± 0.2 of the two Pt inverse residual resistivities, to within mutual uncertainties.

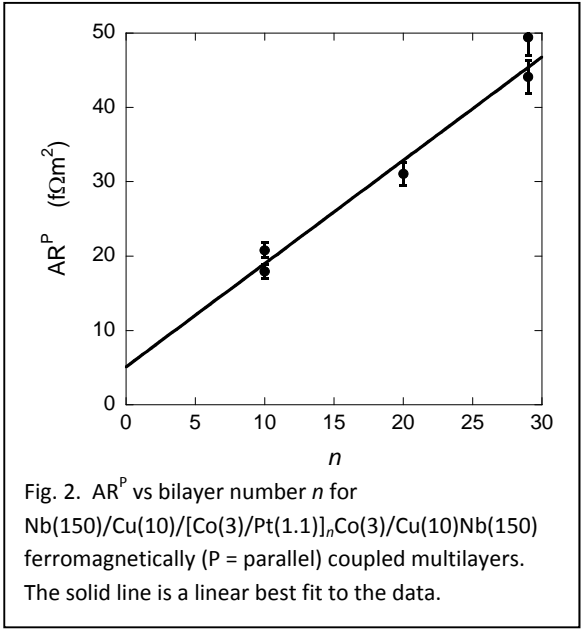

Fig. 2. $AR^P$ vs bilayer number $n$ for Nb(150)/Cu(10)/[Co(3)/Pt(1.1)]$_n$Co(3)/Cu(10)Nb(150) ferromagnetically (P = parallel) coupled multilayers. The solid line is a linear best fit to the data.

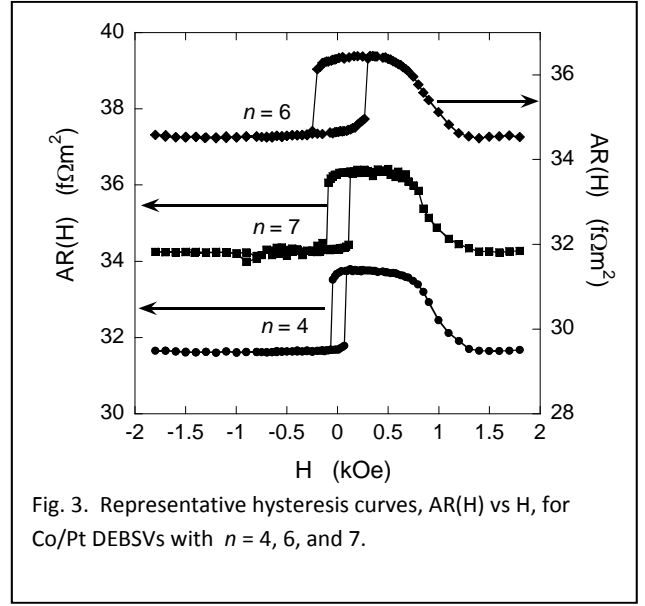

Fig. 3. Representative hysteresis curves, AR(H) vs H, for Co/Pt DEBSVs with $n$ = 4, 6, and 7.

The measurements in the present paper were made on samples sputtered from two Pt pieces in the small gun as just explained. Aside from Pt, van der Pauw measurements of the resistivities of the other metals used in this study agreed to within uncertainties with prior results: Cu (present $\rho_{Cu}$ = 6 ± 1 nΩm vs $\rho_{Cu}$ = 6 nΩm in [20] for Cu also sputtered from a large gun); Py (present $\rho_{Py}$ = 101 ± 10 nΩm vs $\rho_{Py}$ = 123 ± 40 nΩm in [12]); Co (present $\rho_{Co}$ = 46 ± 10 nΩm vs $\rho_{Co}$ = 59 ±10 nΩm [12]). These agreements make us comfortable fixing all of the parameters other than those for Pt and Co/Pt interfaces at the values listed in [12]—except for Cu, where we now use $\rho_{Cu}$ = 6 ± 1 nΩm.

For Pt, we've already explained the values of $\rho_{Pt}$ and $l_{sf}^{Pt}$ that we use. For Co/Pt interfaces, we need to determine three parameters, the $\delta_{Co/Pt}$ that is the focus of the present paper, and the two VF parameters [17] that characterize the properties of Co/Pt interfaces. These can be chosen either as the specific resistances $AR_{Co/Pt}^{\uparrow}$ and $AR_{Co/Pt}^{\downarrow}$ (A = area through which the CPP current flows, and up and down specify that the moments of the conduction electrons are along or opposite to the magnetic moment of the F-layer through which they are passing), or as $AR_{Co/Pt}^{*}$ = ($AR_{Co/Pt}^{\downarrow}$ + $AR_{Co/Pt}^{\uparrow}$)/4 and $\gamma_{CoPt}$ = ($AR_{Co/Pt}^{\downarrow}$ - $AR_{Co/Pt}^{\uparrow}$)/( $AR_{Co/Pt}^{\downarrow}$ + $AR_{Co/Pt}^{\uparrow}$). The present fits were done using $AR_{Co/Pt}^{\uparrow}$ and $AR_{Co/Pt}^{\downarrow}$. But we quote also the values for $AR_{Co/Pt}^{*}$ and $\gamma_{Co/Pt}$, which we compare with our previously published values [21] derived assuming $\delta_{Co/Pt}$ = 0. The three parameters were determined self-consistently from fits to three experimental quantities, as described below. The first experimental quantity was determined from measurements of AR vs $n$ on ferromagnetically coupled [Co(3)/Pt(1.1)]$_n$Co(3) multilayers ($n$ = number of bilayer repeats and thicknesses are given in nm) embedded between 150 nm thick Nb layers. The fit to these data depends only weakly on $\delta_{Co/Pt}$. The second and third experimental quantities were obtained from samples containing more complex multilayers as described in Dassonneville et al. [8]. Each sample contains a ferromagnetically coupled $[Co(3)/Pt(1.1)]_nCo(3)$ multilayer sputtered into the middle of a Permalloy (Py = $Ne_{1-x}Fe_x$ with x ~ 0.2)-based double exchange-biased spin-valve (DEBSV). The sample form is $Nb(150)/Cu(10)/FeMn(8)/Py(6)/Cu(10)/[Co(3)/Pt(1.1)]_n/Co(3)/Cu(10)/Py(6)/FeMn(8)/Cu(10)/Nb(150)$. After fabrication, the sample was heated in vacuum and in a magnetic field ~ 200 Oe to ~ 453K, above the FeMn blocking temperature, and then cooled to room temperature in the presence of the field. This procedure 'pins' the two Py layers next to FeMn to a higher coercive field ($H_c$ ~ 700–800 Oe) than that of the ferromagnetically coupled Co/Pt multilayer ($H_c$ ~ 50–250 Oe). To choose the Pt thickness for ferromagnetic coupling, we sputtered a series of Co/Pt multilayers with Pt thicknesses ranging from 0.8 to 1.6 nm in 0.2 nm steps and measured their magnetizations vs field in a SQUID magnetometer. The samples with $t_{Pt}$ = 1.0 and 1.2 nm gave the lowest $H_c$ and the highest normalized remnent moment. We thus chose $t_{Pt}$ = 1.1 nm for our samples, which is larger than the expected interface thickness ($t_i$ ~ 0.6-0.8 nm)[16]. As noted above, due to interface roughness, ferromagnetic coupling may not be sensitive to the precise Pt thickness. The Co thickness was fixed at $t_{Co}$ = 3 nm to be larger than the expected interface thickness but also so that 8 x $t_{Co}$ = 24 nm was still smaller than the expected spin-diffusion length of $l_{sf}^{Co}$ ~ 60 nm [22,23] in our sputtered Co at 4.2K.

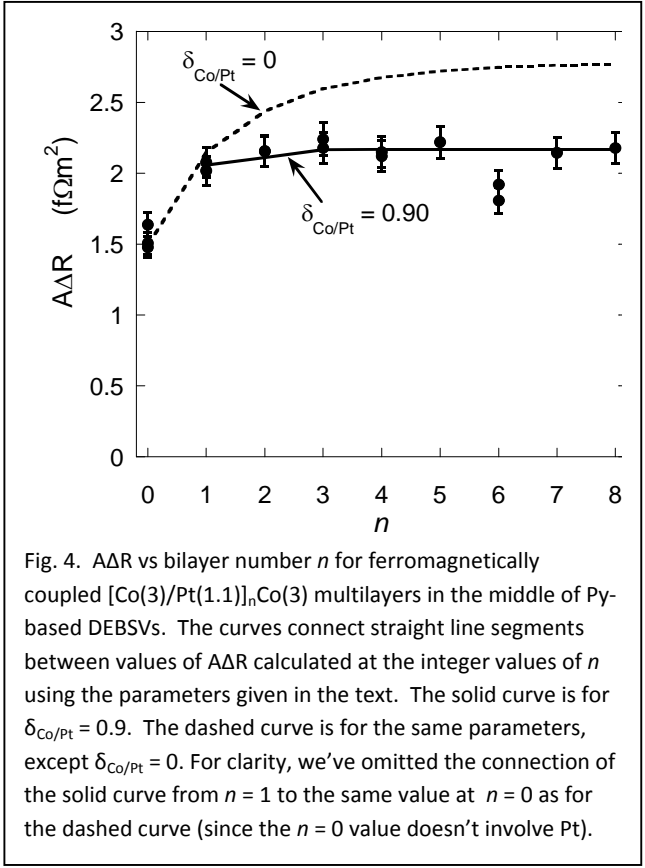

Fig. 4. AΔR vs bilayer number $n$ for ferromagnetically coupled $[Co(3)/Pt(1.1)]_nCo(3)$ multilayers in the middle of Py-based DEBSVs. The curves connect straight line segments between values of AΔR calculated at the integer values of $n$ using the parameters given in the text. The solid curve is for $\delta_{Co/Pt}$ = 0.9. The dashed curve is for the same parameters, except $\delta_{Co/Pt}$ = 0. For clarity, we've omitted the connection of the solid curve from $n$ = 1 to the same value at $n$ = 0 as for the dashed curve (since the $n$ = 0 value doesn't involve Pt).

(2b) Data and Analysis.

As explained above, the data to be analyzed involve two sets of measurements.

The first set is the plot of AR vs $n$ for simple $Nb(150)/Cu(10)/[Co(3)/Pt(1.1)]_nCo(3)/Cu(10)Nb(150)$ multilayers shown in Fig. 2. As expected, because the Co layers are ferromagnetically coupled, each value of AR was independent of applied magnetic field H checked to ± 1.5 kG. The solid line is a least-squares best fit to the five data points. The ordinate intercept, $2AR_{Co/Nb}$ = 5.1 ± 1.5 $f\Omega m^2$ is consistent with our standard value of 6 $f\Omega m^2$ [20,24]. For our analysis, we use the value of AR on this best fit line at $n$ = 29. This choice has the advantages of: (a) using a fit to all of the data; (b) using a value of $n$ where we have two independent data points; and (c) using a value of $n$ large enough so that the effect on the analysis of the exact value of the intercept is minimized.

The second set involves a plot of AΔR vs $n$ for similar ferromagnetically coupled $[Co(3)/Pt(1.1)]_nCo(3)$ multilayers, but now embedded in the middle of a Py-based DEBSV. In this case, AR varies hysteretically with H as illustrated in Fig. 3 for representative samples with three different values of $n$ = 4, 6, and 7. For each sample, AΔR is defined as the difference in AR between the maximum (AP state) and minimum (P state) values of the hysteretic AR data. Note, for use below, that the coercive field of the sample with $n$ = 6 (and also that for the other $n$ = 6 sample—not shown) is much larger (~ 250 Oe) than those for the samples with $n$ = 4 or 7 (~ 100 Oe), as well as those for the remaining samples from $n$ = 0 to 8 not including $n$ = 6. This larger coercive field means that we can have less confidence that the $n$ = 6 data have reached a true AP state. Said another way, the inferred values of AΔR for the n = 6 data are likely to be only lower bounds on the correct AΔR. Unfortunately, we cannot quantify by how much they are too low.

With this last caveat, we plot in Fig. 4 AΔR vs n for our Py-based EBSV data.

Note, firstly, that the two data points for n = 6 lie well below those for the other values of n from 2 to 8. We attribute this difference partly to failure to reach a complete AP state. So we will analyze our data both with and without the two n = 6 points, using the data without as our best estimate, but using the data with to set the uncertainty.

Note, secondly, that the data in Fig. 4 look to have saturated (become constant in value) already by n = 2. This rapid saturation, by itself, indicates that $\delta_{Co/Pt}$ is large, but does not tell us just how large. As the second of our experimental quantities for determining our three unknowns, we use the average value of AΔR from n = 2 to n = 8 (either neglecting or including n = 6 as noted above), and locate this average at n = 7. Averaging the 8 data points should give a more reliable estimate of the saturation value at n = 7 than choosing any single value of n. Interestingly, omitting n = 2 from the average makes no significant difference.

Lastly, we use as our third experimental quantity the average value of AΔR for n = 1, where we have three data points, the lowest two of which are indistinguishable.

Choosing our best prior values and uncertainties for parameters other than those involving Pt, and our best values and uncertainties for the three quantities involving Pt: AR(n = 29) in Fig. 2; AΔR(n = 1) in Fig. 4; and the averaged AΔR(n = 7) in Fig. 4, gives rounded values of $AR^{\downarrow}$ = 2.3 ± 0.5 fΩm², $AR^{\uparrow}$ = 0.7 ± 0.1 fΩm², and $\delta_{Co/Pt}$ = $0.9^{+0.3}_{-0.2}$. Converting the unrounded values to VF parameters gives $AR^*_{Co/Pt}$ = 0.74 ± 0.15 fΩm²; $\gamma_{Co/Pt}$ = 0.53 ± 0.12. If we include the two n = 6 points in the average for AΔR(n = 7), the best averages shift to: $AR^*_{Co/Pt}$ = 0.78 fΩm²; $\gamma_{Co/Pt}$ = 0.57; and $\delta_{Co/Pt}$ = 1.2, all within the uncertainties specified without the n = 6 points. But now the upward uncertainty in $\delta_{Co/Pt}$ also increases to where, if the n = 6 data are taken at face value, $\delta_{Co/Pt}$ as large as 2.0 cannot be absolutely ruled out. Given that the n = 6 data are probably only lower bounds, we choose as our final best estimates and uncertainties: $AR^*_{Co/Pt}$ = 0.74 ± 0.15 fΩm²; $\gamma_{Co/Pt}$ = 0.53 ± 0.12 and $\delta_{Co/Pt}$ = $0.9^{+0.5}_{-0.2}$. These values of $AR^*_{Co/Pt}$ and ; $\gamma_{Co/Pt}$ overlap with those in [21], $AR^*_{Co/Pt}$ = 0.85 ± 0.13 and $\gamma_{Co/Pt}$ = 0.38 ± 0.06, derived with different kinds of multilayers and neglecting spin-flipping at the Co/Pt interface, i.e., assuming $\delta_{Co/Pt}$ = 0.

3. Summary and Conclusions.

This paper contains three main results.

(1) The tendency of the data for deposited layers of Pt in Fig. 1 to fall roughly along a single straight line demonstrates that the spin-diffusion length in Pt, $l^{Pt}_{sf}$, a measure of spin-flip scattering, is not intrinsic to Pt, but rather is approximately proportional to the inverse resistivity, $1/\rho_{Pt}$, a measure of total scattering. The variations in the data for similar values of $1/\rho_{Pt}$ indicate that this proportionality is not perfect, with $l^{Pt}_{sf}$ differing somewhat for different scatterers.

(2) We derived new values of the Valet-Fert parameters for Co/Pt interfaces: $AR^*_{Co/Pt}$ = 0.74 ± 0.15 fΩm² for the enhanced interface specific resistance, and $\gamma_{Co/Pt}$ = 0.53 ± 0.12 for the interface scattering asymmetry. Both values are similar to those found for other Co/N (N = non-magnetic metal) interfaces, [20,25] and overlap with our previous values derived assuming $\delta_{Co/Pt}$ = 0 [21].

(3) We found an interface spin-flipping parameter for Co/Pt, $\delta_{Co/Pt}$ = $0.9^{+0.5}_{-0.2}$, that is much larger than those ~ 0.2-0.35 found for five other pairs involving one or two ferromagnetic (F) metals. While this particular value of $\delta_{Co/Pt}$ = $0.9^{+0.5}_{-0.2}$ is derived for intermixed interfaces of sputtered Co and Pt, $\delta_{F/Pt}$ for any other F-metal and any interfacial structure with Pt is probably large enough so that one must correct for significant degradation of a polarized current crossing such an interface. But to know by exactly how much the current will degrade in a specific case other than sputtered Co/Pt, one needs to know two things. Is our large value for sputtered Co/Pt due primarily to a large spin-orbit difference, or is it produced partly by a proximity effect that turns Pt magnetic over a small distance from an F/Pt boundary [14]? And, could it be produced by a perfect interface, or is interfacial intermixing important? We hope that our result will stimulate theorists to help answer these questions.